\documentclass[runningheads]{llncs}
\usepackage{graphicx,hyperref, enumerate}
\usepackage{listings,mdframed,xcolor,booktabs}
\usepackage{tikz}
\pgfdeclarelayer{background}
\pgfsetlayers{background,main}
\usepackage{algorithm, algorithmic, bm}
\usetikzlibrary{mindmap,shadows,calc,shadows.blur}
\usetikzlibrary{patterns,arrows,positioning,fit,petri,mindmap,trees,intersections,cd}

\lstdefinelanguage{sparqlB}{
    morecomment=[l]{\#},
    language=SQL,
    morekeywords={SELECT},      
    commentstyle=\itshape\color{blue},
    sensitive=true,
}

\definecolor{dkblue}{rgb}{0,0.1,0.5}
\definecolor{lightblue}{rgb}{0,0.5,0.5}
\definecolor{dkgreen}{rgb}{0,0.4,0}
\definecolor{dk2green}{rgb}{0.4,0,0}
\definecolor{dkviolet}{rgb}{0.6,0,0.8}
\definecolor{mantra}{rgb}{0.2,0.6,0.2}
\definecolor{gotcha}{rgb}{0.8,0.2,0}
\definecolor{ocre}{RGB}{243,102,25} 
\definecolor{mint}{HTML}{44BB99}
\newcommand{\Co}[1]{\mbox{\lstinline `#1`}}
\newcommand{\dtkg}{DTKG}

\definecolor{blueTOL}{HTML}{004488}
\definecolor{yellowTOL}{HTML}{DDAA33}
\definecolor{redTOL}{HTML}{BB5566}
\definecolor{wine}{HTML}{882255}
\definecolor{grey}{HTML}{BBBBBB}

\begin{document}

\lstnewenvironment{sparqlinonly}{
\vspace{0.5\baselineskip}%
\noindent
\minipage{1.2\textwidth}%
\mdframed[skipabove=0pt,
skipbelow=2pt,
rightline=false,
leftline=true,
topline=false,
bottomline=false,
linecolor=mint,
innerleftmargin=5pt,
innerrightmargin=5pt,
innertopmargin=0pt,
leftmargin=0cm,
rightmargin=0cm,
linewidth=4pt,
innerbottommargin=0pt]\lstset{
numbers=left,numberstyle=\tiny,xleftmargin=5mm,
language=sparql,basicstyle=\footnotesize,
morekeywords={SELECT, REASON, OVER}}%
}
{\endmdframed%
\endminipage%
\vspace{0.5\baselineskip}
}

\lstnewenvironment{sparqlin}{
\vspace{0.125\baselineskip}%
\noindent
\minipage{\textwidth}%
\mdframed[skipabove=0pt,
skipbelow=2pt,
rightline=false,
leftline=true,
topline=false,
bottomline=false,
linecolor=mint,
innerleftmargin=5pt,
innerrightmargin=5pt,
innertopmargin=-3pt,
leftmargin=0cm,
rightmargin=0cm,
linewidth=4pt,
innerbottommargin=-3pt]\lstset{
language=sparql,basicstyle=\footnotesize,
morekeywords={SELECT, REASON, OVER}}%
}
{\endmdframed%
\endminipage%
\vspace{0.0625\baselineskip}
}

\lstnewenvironment{sparqlout}{
\vspace{0.0625\baselineskip}%
\noindent
\minipage{\textwidth}%
\mdframed[skipabove=0pt,
skipbelow=7pt,
rightline=false,
leftline=true,
topline=false,
bottomline=false,
linecolor=gray,
backgroundcolor=black!5,
innerleftmargin=5pt,
innerrightmargin=5pt,
innertopmargin=-3pt,
leftmargin=0cm,
rightmargin=0cm,
linewidth=4pt,
innerbottommargin=-3pt]\lstset{language=sparql,basicstyle=\footnotesize,
morekeywords={SELECT, REASON, OVER}}%
}
{\endmdframed%
 \endminipage%
 \vspace{0.125\baselineskip}
}

\lstnewenvironment{coqin}{
\vspace{0.125\baselineskip}%
\noindent
\minipage{\textwidth}%
\mdframed[skipabove=0pt,
skipbelow=7pt,
rightline=false,
leftline=true,
topline=false,
bottomline=false,
linecolor=ocre,
innerleftmargin=5pt,
innerrightmargin=5pt,
innertopmargin=-3pt,
leftmargin=0cm,
rightmargin=0cm,
linewidth=4pt,
innerbottommargin=-3pt]%
}
{\endmdframed%
 \endminipage%
 \vspace{0.0625\baselineskip}
}

\lstnewenvironment{coqout}{
\vspace{0.0625\baselineskip}%
\noindent
\minipage{\textwidth}%
\mdframed[skipabove=0pt,
skipbelow=7pt,
rightline=false,
leftline=true,
topline=false,
bottomline=false,
linecolor=gray,
backgroundcolor=black!5,
innerleftmargin=5pt,
innerrightmargin=5pt,
innertopmargin=-3pt,
leftmargin=0cm,
rightmargin=0cm,
linewidth=4pt,
innerbottommargin=-3pt]%
}
{\endmdframed%
 \endminipage%
 \vspace{0.125\baselineskip}
}

\lstnewenvironment{coqinonly}{
\vspace{0.25\baselineskip}%
\noindent
\minipage{\textwidth}%
\mdframed[skipabove=0pt,
skipbelow=7pt,
rightline=false,
leftline=true,
topline=false,
bottomline=false,
linecolor=ocre,
innerleftmargin=5pt,
innerrightmargin=5pt,
innertopmargin=-3pt,
leftmargin=0cm,
rightmargin=0cm,
linewidth=4pt,
innerbottommargin=-3pt]%
}
{\endmdframed%
 \endminipage%
 \vspace{0.25\baselineskip}
}

\lstnewenvironment{coqinnum}{
\vspace{0.5\baselineskip}%
\noindent
\minipage{\textwidth}%
\mdframed[skipabove=0pt,
skipbelow=7pt,
rightline=false,
leftline=true,
topline=false,
bottomline=false,
linecolor=ocre,
innerleftmargin=5pt,
innerrightmargin=5pt,
innertopmargin=0pt,
leftmargin=0cm,
rightmargin=0cm,
linewidth=4pt,
innerbottommargin=0pt]
\lstset{numbers=left,numberstyle=\tiny,xleftmargin=5mm}}%
{\endmdframed%
 \endminipage%
 \vspace{0.5\baselineskip}
}

\title{Dependently Typed Knowledge Graphs\thanks{Supported by NVIDIA AI Technology Centre.}}
%
%
\author{Zhangsheng Lai\inst{1,2} 
\and
Aik Beng Ng\inst{1,2}
\and
Liang Ze Wong\inst{3}
\\ \and
Simon See \inst{1} \and
Shaowei Lin \inst{2}}
\authorrunning{Z. Lai, A. B. Ng et al.}
%
\institute{NVIDIA AI Technology Centre, Singapore\\
\email{\{zlai,aikbengn,ssee\}@nvidia.com}
\and
Singapore University of Technology and Design, Singapore
\email{shaowei\_lin@sutd.edu.sg}
 \and
Institute of High Performance Computing, Singapore\\
\email{wonglz@ihpc.a-star.edu.sg}}
\maketitle              
\begin{abstract}
Reasoning over knowledge graphs is traditionally built upon a hierarchy of languages in the Semantic Web Stack. 
Starting from the Resource Description Framework (RDF) for knowledge graphs, more advanced constructs have been introduced through various syntax extensions to add reasoning capabilities to knowledge graphs. 
In this paper, we show how standardized semantic web technologies (RDF and its query language SPARQL) can be reproduced in a unified manner with dependent type theory. 
In addition to providing the basic functionalities of knowledge graphs, dependent types add expressiveness in encoding both entities and queries, explainability in answers to queries through witnesses, and compositionality and automation in the construction of witnesses.
Using the Coq proof assistant, we demonstrate how to build and query dependently typed knowledge graphs as a proof of concept for future works in this direction.
\keywords{dependent type theory \and knowledge graphs \and reasoning.}
\end{abstract}

\section{Introduction}
The encoding of information in machine-understandable formats in knowledge graphs, formalized through linked data protocols, has led to its ubiquity and use in various tasks such as public web search, private knowledge management, and question answering \cite{Cui:2017:KLQ:3055540.3055549,DBLP:journals/corr/abs-1811-04540,gkg12,Noy2019}. Knowledge graphs store everything from static general facts \cite{LOD} to dynamic sensor data \cite{le2016graph} and from functions and algorithms \cite{buswell2004open} to rules and theorems \cite{kifer2013rif}. The diversity in the kinds of objects which are represented in knowledge graphs is summarized in the Semantic Web Stack \cite{semanticwebstack}. The stack includes both basic languages, such as the Resource Description Framework (RDF) \cite{Lassila1999} for describing data objects in the graph, and higher-order languages, such as the Web Ontological Language (OWL) \cite{hitzler2009owl} for describing logical relationships between concepts. For machines to exploit the spectrum of information encapsulated in knowledge graphs for reasoning, they will need a logical theory in which these objects can be understood and manipulated \cite{CHEN2020112948}.

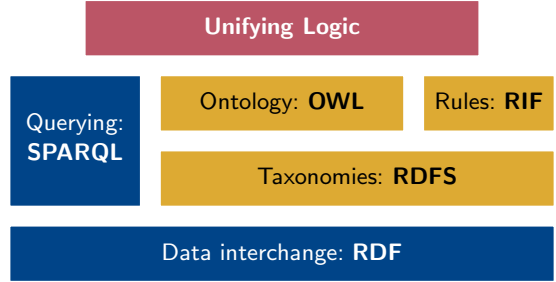
\begin{figure}
\centering
\begin{tikzpicture}[]
\tikzstyle{label}=[align=center]
\tikzstyle{layer}=[draw, label, anchor=south west, minimum height=.75cm, minimum width=1cm,
]
  \draw[draw=white,thick] (.5,2) node (rdf) [layer, minimum width=7.25cm,fill=blueTOL]           {{\color{white}\textsf{Data interchange: \textbf{RDF}}}\strut};
  \draw[draw=white,thick] (.5,3) node   (query)   [layer, minimum width=1.75cm, minimum height=1.75cm, text width=1.5cm,fill=blueTOL] {{\color{white}\textsf{Querying:}\\ \textsf{\textbf{SPARQL}}}\strut};

  \draw[draw=white,thick] (2.5,3) node  (rdfs)    [layer, minimum width=5.25cm, fill = yellowTOL]           {\textsf{Taxonomies: \textbf{RDFS}}\strut};
  \draw[draw=white,thick] (2.5,4) node  (owl)     [layer, minimum width=3.25cm, fill = yellowTOL]           {\textsf{Ontology: \textbf{OWL}}\strut};
  \draw[draw=white,thick] (6,4) node    (rif)     [layer, minimum width=1.75cm, fill = yellowTOL ] {\textsf{Rules: \textbf{RIF}}\strut};
  \draw[draw=white,thick] (1.5,5) node  (logic)   [layer, minimum width=5.25cm, fill = redTOL]           {{\color{white}\textsf{\textbf{Unifying Logic}}}\strut};


\end{tikzpicture}
\caption{Part of the Semantic Web Stack, adapated from \cite{semanticwebstack}.  In this work, we reproduce the functionalities of RDF and SPARQL (boxes in blue) using dependent type theory.}
\label{fig:SWS}
\end{figure}

A crucial challenge to be tackled by such a logical theory is \textit{provenance}, the tracking of data used in computing a new piece of information for the purpose of assessing the quality and trustworthiness of the computation \cite{moreau2010foundations}. 
For example, if a knowledge graph contains the claim that ``Barack is the father of Malia'', it may be unclear if the claim is directly witnessed by some evidence such as a birth certificate, or indirectly witnessed by a composite of the claims ``Barack is the husband of Michelle'' and ``Michelle is the mother of Malia'' from the same graph. 
In many critical applications, we often need to go beyond the list of witnesses used in a composite to an explicit description of how that composite was constructed. 
We argue that a good solution to this challenge is a symbolic calculus of witnesses, which is the principle behind constructive mathematics \cite{bertot2013interactive}. 
Moreover, we propose using dependent type theory for this purpose, specifically in the form of the Calculus of Inductive Constructions as implemented by the software Coq \cite{COQ}. 

Our starting point is the Curry-Howard correspondence, which encodes theorems and proofs in mathematical logic as types and terms in dependent type theory \cite{sorensen2006lectures}. 
Given a type, there are powerful machine-tactics for building terms belonging to that type, due to the constructive nature of dependent type theory \cite{milner1984use}. 
As a result, the Curry-Howard correspondence provides strategies for proving mathematical theorems with help from a machine, which is the guiding philosophy behind successful proof-assistants such as Coq, Agda and Lean \cite{COQ,AGDA,LEAN}. 
We extend the correspondence by viewing queries on knowledge graphs as types and their witnesses (answers with proof) as terms in dependent type theory. 
This paper explores how the extended view provides not just a language for encoding queries and their witnesses, but also powerful machine-tactics for deriving witnesses for a given query.

Under this \emph{queries-as-types} approach, we regard Uniform Resource Identifiers (URIs) as primitive constants (constructors) or defined constants (definitions) in the syntax of the type theory \cite{nordstrom2000martin}. 
These URIs constitute a vocabulary that is shared across all the entities and relations in the knowledge graph, and are the building blocks for our type-theoretic universe.
In this paper, we demonstrate how RDF triples and SPARQL queries are jointly represented in this universe. 

In Section \ref{sec:repKG}, we focus on representations of knowledge graphs in the Resource Description Framework (RDF) and in dependent type theory.
RDF triples lay out type information for each entity as well as relations between the entities \cite{Lassila1999}, and are the lingua franca of the Semantic Web.
To encode RDF graphs in dependent type theory, we group the vertices (e.g. entities of the form ``$x$ is a person'') and edges (e.g. relations of the form ``$x$ is the father of $y$'') of the knowledge graph into families, and view each family as a type and its members as terms of that type. 
We call the resulting encoding a \emph{dependently typed knowledge graph} (DTKG), and formalize this encoding in Coq.

In Section \ref{sec:queryKG}, we turn to the querying of knowledge graphs.
For RDF graphs, this is carried in the SPARQL Protocol and RDF Query Language. 
For DTKGs, SPARQL queries (e.g. ``Who is the father of Malia?'') \cite{perez2006semantics} will be regarded as record types where the primary fields (e.g. father) coerce to some parent type (e.g. people) and where the secondary fields store proofs of conditions (e.g. ``is the father of Malia'') satisfied by the primary field. 
These record types may be thought of as a form of subtyping, although traditional subtypes in type theory additionally impose a restriction that the secondary fields be defined by boolean predicates \cite{mahboubi2017mathematical}. 
We describe the encoding of both direct and composite queries as record types. 

To construct witnesses for a given query, we call on tactics in proof-assistants to decompose the query into simpler subqueries \cite{milner1984use}. 
Previously-constructed witnesses for the subqueries will be exploited during this process, and the search for a suitable decomposition can be automated to a large extent \cite[VII]{pierce2010software}. 
Such automation is possible because both atomic entities and composite witnesses are decorated with rich type information. 

In summary, the advantages of using dependent type theory as a underlying logic for reasoning over knowledge graphs are its expressiveness in representing both entities and queries, the provenance and explainability afforded by witnesses constructed for the queries, and the degree of compositionality and automation available for the construction of witnesses. 
As far as we know, our work is the first in exploring how knowledge graphs can be encoded within dependent type theory for reasoning.

\section{Representations of Knowledge Graphs} \label{sec:repKG}

\begin{figure}[h] 
\centering
\begin{tikzpicture}[scale = 0.75, node distance=\layersep,>=latex] 
    \tikzstyle{neuron}=[circle,fill=black!25,minimum size=15pt,inner sep=0pt];
    \tikzstyle{unit}=[neuron, fill=blueTOL,
    ];
    \tikzstyle{spike}=[neuron, fill=blue!50];
 \def \radius {1.5cm}
\begin{scope}[xshift=2.5cm]
 \def \n {3}
    \node[unit,draw=white,thick] (t1) at (-180:\radius) {};
    \node[unit,draw=white,thick] (t2) at (-60:\radius) {};
    \node[unit,draw=white,thick] (t3) at (60:\radius) {};
   \draw[->] (t1) -- (t2);
   \draw[->] (t2) edge node[right] {$\mathsf{Relation/Edge}$} (t3);
   \draw[->] (t3) -- (t1);
    \node (l) at (27.5:2.85) {$\mathsf{Entity/Vertex}$};   

\end{scope}

\begin{scope}[xshift=-2.5cm]
 \def \n {4}
 \foreach \s in {1,...,\n}
  \node[unit,draw=white,thick] (\s) at ({360/\n * (\s - 1) - 180}:\radius) {};
  \draw[->] (1) -- (2);
  \draw[->] (2) -- (3);
  \draw[->] (3) -- (4);
  \draw[->] (4) -- (1);
\end{scope}
   \draw[->] (3) edge node (e) {} (t1) ;

\node at (-2,-3) (s) {$\mathsf{subject}$};
\node at (0,-3) (p) {$\mathsf{predicate}$};
\node at (2,-3) (o) {$\mathsf{object}$};
\draw[->,ultra thick] (3) -- (s);
\draw[->,ultra thick] (e) -- (p);
\draw[->,ultra thick] (t1) -- (o);
\node at (-4,-4.5) (s1) {\Co{Sasha}};
\node at (0,-4.5) (p1) {\Co{hasFather}};
\node at (5.5,-4.5) (o1) {\Co{Barack}};
\draw[->,ultra thick, dashed] (s) -- (s1);
\draw[->,ultra thick, dashed] (p) -- (p1);
\draw[->,ultra thick, dashed] (o) -- (o1);
\end{tikzpicture}
\caption{A knowledge graph and its components}\label{fig:structureofkg}
\end{figure}
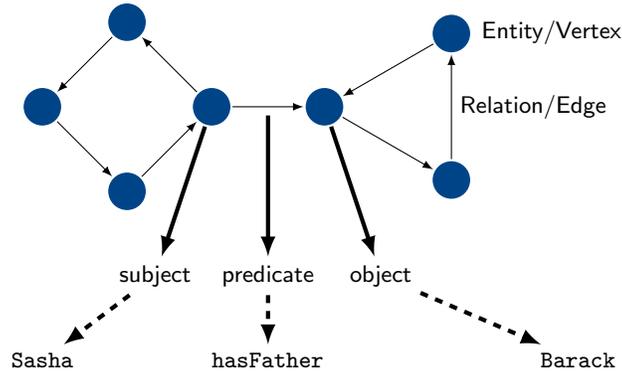

Knowledge graphs such as DBpedia \cite{Auer2007dbpedia}, NELL \cite{Carlson2010nell} and YAGO \cite{Suchanek2007yago} provide structured data and factual information, and have been widely used in applications like question answering, recommender systems and search engines \cite{Cui:2017:KLQ:3055540.3055549,DBLP:journals/corr/abs-1811-04540,gkg12,Noy2019}. A knowledge graph can be expressed as a collection of triples, each consisting of a subject, a predicate and an object:
\begin{equation}
\label{eq:KG-def}
\mathcal{KG} := \{\Co{(subj, pred, obj)}\mid \Co{subj, obj} \in \mathcal{E}, \Co{pred} \in \mathcal{R}\}
\label{eq:kg}
\end{equation}
where $\mathcal{E}$ is a set of entities, and $\mathcal{R}$ is a set of relations or predicates. 
In graph terminology, $\mathcal{E}$ is the set of vertices and $\mathcal{KG}$ is the set of directed edges. 
Each triple in $\mathcal{KG}$ is a directed edge whose source (resp. target) is the subject (resp. object) of the predicate.
Fig \ref{fig:structureofkg} illustrates the structure of a knowledge graph and the terminology for describing its components.

Note that the same \emph{relation} (e.g. \Co{hasFather}) can occur in multiple \emph{edges}.
For example, a knowledge graph could contain both
\Co{(Sasha, hasFather, Barack)} and \Co{(Malia, hasFather, Barack)}.

In this section, we describe the encoding of knowledge graphs as RDF graphs as well as \dtkg{}s. We provide an algorithm to convert RDF graphs into DTKGs and discuss the advantages of DTKGs over RDF graphs. 

\subsection{RDF knowledge graphs}

The Resource Description Framework (RDF) \cite{Lassila1999} is a popular framework for representing knowledge in graph structures. The RDF is a graph-based data model structured as a set of triples, shown in (\ref{eq:kg}). A set of such triples is called an \emph{RDF graph}, with the components of the triple (i.e. the elements of $\mathcal{E}$ and $\mathcal{R}$) represented using Uniform Resource Identifiers (URIs). 

The RDF format, however, imposes little to no constraints on the triples that can be created. It provides a basic format to express relationships between resources and is always used in combination with the RDF Schema (RDFS) \cite{brickley2014rdfs}, which exploits the capabilities of the RDF layer to describe the semantic information of these resources. 
Although RDFS allows us to prescribe the class information of the subject and object of a predicate, these specifications are still not restrictions on the predicate. Instead, they merely provide guidance to help discover possible errors, suggest appropriate values in an interactive editor or infer additional information in a reasoning application. 

RDF graphs can be serialized in several different formats. In this paper, we will use the more readable Turtle \cite{turtle} format (e.g. in Fig \ref{fig:obamardf}).






\begin{example}
\begin{figure}[t]
\centering
\begin{tikzpicture}[scale = 1, node distance=\layersep,>=stealth]
    \tikzstyle{neuron}=[circle,fill=black!25,minimum size=17pt,inner sep=0pt];
    \tikzstyle{unit}=[neuron, fill=red!50,thick];
    \tikzstyle{babi}=[rectangle,fill=blueTOL,inner sep=3pt,minimum size=10pt,rounded corners=1mm,
    ];
    \tikzstyle{meta}=[rectangle,fill=redTOL,inner sep=3pt,minimum size=10pt,rounded corners=1mm,
    ];

\node[babi,draw=white,thick] (f) at (-1.5,0) {{\color{white}\textsf{\scriptsize barack}}};
\node[babi,draw=white,thick] (d) at (4.75,0) {{\color{white}\textsf{\scriptsize malia}}};
\node[babi,draw=white,thick] (s) at (-4.5,0) {{\color{white}\textsf{\scriptsize sasha}}};
\node[babi,draw=white,thick] (m) at (1.5,0) {{\color{white}\textsf{\scriptsize michelle}}};

\path (s) edge [->, thick]   node[above,sloped] (nf) {\scriptsize$\mathsf{father}$} (f);
\path (d) edge [->, thick]    node[above ,sloped] (nm) {\scriptsize$\mathsf{mother}$} (m);
\path (m) edge [->, thick]      node[above ,sloped] (nh) {\scriptsize$\mathsf{husband}$} (f);
\end{tikzpicture}
\caption{Knowledge graph of the Obama family}
\label{fig:obamagraph-simple}
\end{figure}
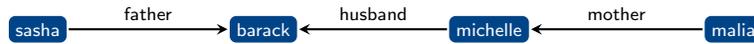

As a running example, consider the knowledge graph in Fig \ref{fig:obamagraph-simple} which displays some information about the Obama family.
The RDF format allows us to encode this knowledge graph as an RDF graph, while RDFS allows us to specify that the entities \Co{Barack, Michelle, Malia, Sasha} are of class \Co{Person} (as opposed to \Co{Place}, for example).
The code in Fig \ref{fig:obamardf} shows the encoding of this knowledge graph using RDF and RDFS.

\begin{figure}
\begin{sparqlinonly}
@PREFIX ex: <http://example.org/>
@PREFIX rdf: <http://xmlns.com/foaf/0.1/>
@PREFIX rdfs: <http://www.w3.org/2000/01/rdf-schema#>

ex:Person rdf:type rdfs:Class.

ex:father rdf:type rdf:Property; rdfs:domain ex:Person; rdfs:range ex:Person.
ex:mother rdf:type rdf:Property; rdfs:domain ex:Person; rdfs:range ex:Person.
ex:husband rdf:type rdf:Property; rdfs:domain ex:Person; rdfs:range ex:Person.

ex:Barack rdf:type ex:Person.
ex:Michelle rdf:type ex:Person; ex:husband ex:Barack.
ex:Malia rdf:type ex:Person; ex:mother ex:Michelle.
ex:Sasha rdf:type ex:Person; ex:father ex:Barack.
\end{sparqlinonly}

\caption{RDF data in Turtle format for the knowledge graph in Fig \ref{fig:obamagraph-simple}.}
\label{fig:obamardf}
\end{figure}

Lines 1-3 contain user-defined prefixes that help to shorten the code and improve readability.
Line 5 of Fig \ref{fig:obamardf} declares a class called \Co{Person}. 
Lines 7-9 define the predicates (or properties) \Co{father, mother, husband} whose domains and ranges are expected to be of class \Co{Person}.
Lines 11-14 define the entities \Co{Barack, Michelle}, etc. which are of class \Co{Person}, along with their relevant predicates.
For example, the second half of line 12 declares that the \Co{husband} of \Co{Michelle} is \Co{Barack}.

Note that class assertions like \Co{ex:Person rdf:type rdfs:Class} (line 5) are also treated as edges of the RDF graph.
The RDF graph generated by the code in Fig \ref{fig:obamardf} would thus look like the graph in Fig \ref{fig:obamagraph}.
\end{example}

\begin{figure}
\centering
\begin{tikzpicture}[scale = 1, node distance=\layersep,>=stealth]
    \tikzstyle{neuron}=[circle,fill=black!25,minimum size=17pt,inner sep=0pt];
    \tikzstyle{unit}=[neuron, fill=red!50,thick];
    \tikzstyle{babi}=[rectangle,fill=blueTOL,inner sep=3pt,minimum size=10pt,rounded corners=1mm,
    ];
    \tikzstyle{meta}=[rectangle,fill=redTOL,inner sep=3pt,minimum size=10pt,rounded corners=1mm,
    ];
 \def \radius {1.75cm}
\node[babi,draw=white,thick] (f) at (180: \radius) {{\color{white}\textsf{\scriptsize ex:Barack}}};
\node[babi,draw=white,thick] (d) at (90: \radius) {{\color{white}\textsf{\scriptsize ex:Malia}}};
\node[babi,draw=white,thick] (s) at (270: \radius) {{\color{white}\textsf{\scriptsize ex:Sasha}}};
\node[babi,draw=white,thick] (m) at (0: \radius) {{\color{white}\textsf{\scriptsize ex:Michelle}}};

\node[meta,draw=white,thick] (p) at (5,0) {{\color{white}\textsf{\scriptsize ex:Person}}};

\node[meta,draw=white,thick] (c) at (6,1.75) {{\color{white}\textsf{\scriptsize rdfs:Class}}};
\node[meta,draw=white,thick] (ppty) at (-4,0) {{\color{white}\textsf{\scriptsize rdf:Property}}};

\path (s) edge [->, thick]   node[below,sloped] (nf) {\scriptsize$\mathsf{ex\!:\!father}$} (f);
\path (d) edge [->, thick]    node[below ,sloped] (nm) {\scriptsize$\mathsf{ex\!:\!mother}$} (m);
\path (m) edge [->, thick]      node[above ,sloped] (nh) {\scriptsize$\mathsf{ex\!:\!husband}$} (f);

\path (nf) edge [->,thick, bend left=20, draw=grey]   node[above,sloped] {{\color{grey}\scriptsize$\mathsf{rdf\!:\!type}$}} (ppty);
\path (nm) edge [->,thick, bend right=40, draw=grey]   node[above,sloped] {{\color{grey}\scriptsize$\mathsf{rdf\!:\!type}$}} (ppty);
\path (nh) edge [->,thick, bend right=20, draw=grey]   node[above,sloped] {{\color{grey}\scriptsize$\mathsf{rdf\!:\!type}$}} (ppty);

\path (nm) edge [->,thick, bend left=10, draw=grey]   node[pos=0.3,above,sloped] {{\color{grey}\scriptsize$\mathsf{rdfs\!:\!domain}$}} (p);
\path (nh) edge [->,thick, bend right=20, draw=grey]   node[pos=0.2,below,sloped] {{\color{grey}\scriptsize$\mathsf{rdfs\!:\!domain}$}} (p);
\path (nf) edge [->,thick, bend right=30, draw=grey]   node[pos=0.2,above,sloped] {{\color{grey}\scriptsize$\mathsf{rdfs\!:\!domain}$}} (p);

\path (f) edge [->, thick, bend right=30, draw=grey] node[pos=0.2,above,sloped] {{\color{grey}\scriptsize$\mathsf{rdf\!:\!type}$}} (p);
\path (d) edge [->, thick, bend left=15, draw=grey] node[above,sloped] {{\color{grey}\scriptsize$\mathsf{rdf\!:\!type}$}} (p);
\path (s) edge [->, thick, bend right=30, draw=grey] node[pos=0.7,below,sloped] {{\color{grey}\scriptsize$\mathsf{rdf\!:\!type}$}} (p);
\path (m) edge [->, thick, draw=grey] node[above,sloped] {{\color{grey}\scriptsize$\mathsf{rdf\!:\!type}$}} (p);
\path (p) edge [->, thick, draw=grey] node[above,sloped] {{\color{grey}\scriptsize$\mathsf{rdf\!:\!type}$}} (c);
\end{tikzpicture}
\caption{RDF graph generated by the code in Fig \ref{fig:obamardf}. Semantic information is described by the resources in red. We only indicate the domain of the properties for brevity.
}\label{fig:obamagraph}
\end{figure}
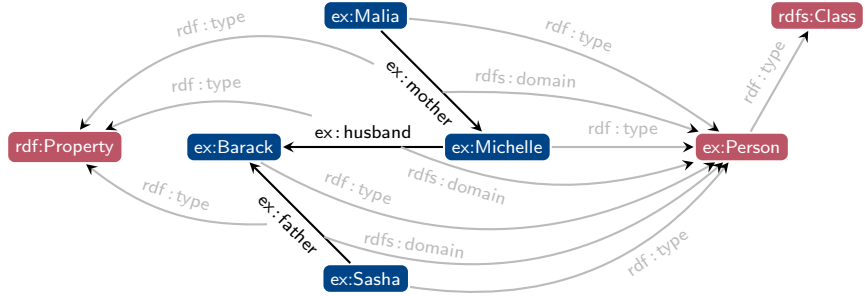

As noted earlier, although this example declares the domain and range of \Co{father} to be of class \Co{Person}, this is not enforced by the system.
If we had another class \Co{Place} and an entity \Co{Chicago} of class \Co{Place}, we could very well have included \Co{ex:Chicago ex:father ex:Barack} in our RDF graph, with no errors being raised.

\subsection{Dependently typed knowledge graphs}\label{sec:DTKG}

A \emph{dependently typed knowledge graph} (DTKG) is a knowledge graph endowed with type information. Entities are terms of \emph{enumerated types}, while directed edges between entities are terms of \emph{dependent types}. 
This is best illustrated by example.

\begin{example}
In Fig \ref{fig:obamacoq}, we encode the RDF graph from Fig \ref{fig:obamardf} as a DTKG, using the Coq proof assistant.
The first line uses the \Co{Inductive} command to define a type called \Co{Person}, and immediately populates it with the \emph{terms} \Co{barack}, \Co{michelle}, \Co{malia} and \Co{sasha}.
This is an \emph{enumerated} type because it is defined by simply enumerating all terms of the type.

\begin{figure}
	\begin{coqinnum}
	Inductive Person := barack | michelle | malia | sasha.

	Inductive HusbandOf : Person -> Person -> Type :=
	  | witness_hmb : HusbandOf michelle barack.
	Inductive MotherOf : Person -> Person -> Type :=
	  | witness_mmm : MotherOf malia michelle.
	Inductive FatherOf : Person -> Person -> Type :=
	  | witness_fsb : FatherOf sasha barack.	  \end{coqinnum}

	\caption{The RDF graph of Fig \ref{fig:obamardf} expressed as a dependently typed knowledge graph in the Coq proof assistant.}
	\label{fig:obamacoq}
\end{figure}

By contrast, the types \Co{HusbandOf, MotherOf} and \Co{FatherOf} (lines 3-8) are \emph{dependent} types whose definitions depend on previously defined types such as \Co{Person}.
Line 3 states that for every pair of terms \Co{x} and \Co{y} in type \Co{Person}, we define a type \Co{HusbandOf x y}. 
Line 4 then populates the type \Co{HusbandOf michelle barack} with the term \Co{witness_hmb}, while all other types of the form \Co{HusbandOf x y} remain empty.

We may interpret relations represented by inhabited types (such as \Co{HusbandOf} \Co{michelle} \Co{barack}) as being true, while relations represented by uninhabited types (such as \Co{HusbandOf malia sasha}) are false.
We call the terms of these types \emph{witnesses} as they represent witnesses of the truth of the relations that they belong to.
As we shall see in Section \ref{ss:queriesasrecords}, these witnesses will be used to provide explainability for the answers to queries.

Note that in this example, there are no witnesses for the relation \Co{FatherOf} \Co{malia} \Co{barack}, even though we might be able to deduce this relation from the relations \Co{MotherOf malia michelle} and \Co{HusbandOf michelle barack}. 
We will return to this issue in Section \ref{ss:compq}.

\end{example}

\subsection{Advantages of DTKGs over RDF graphs}\label{ss:adv-dtkg-rdf}
The biggest advantage of DTKGs over RDF graphs is that the type specifications of entities in DTKGs are enforced restrictions, whereas the class specifications in RDF graphs are not.
Thus, in Fig \ref{fig:obamacoq}, the subject of the relation \Co{HusbandOf} has to be of type \Co{Person}.
If we had another type \Co{Place} containing a term \Co{chicago}, a term of the form \Co{witness_hcb : HusbandOf chicago barack} would raise a type error at compile-time.

A less critical difference is that a relation in a DTKG may contain multiple witnesses to its veracity.
For example, the relation \Co{Fatherof sasha barack} could contain the terms \Co{witness_BC} and \Co{witness_DNA}, representing a birth certificate and a DNA test, respectively.
By contrast, a relation in an RDF graph is either true or false, represented by the presence or absence of an edge.

The advantages listed above are only with regards to the \emph{representation} of a knowledge graphs as an RDF graph or a DTKG.
Further advantages with regards to \emph{querying} these graphs will be mentioned in the Section \ref{ss:featuresdtkg}. 

\subsection{Converting an RDF graph to a DTKG}


The process of converting a knowledge graph into a \dtkg{} involves converting the class and property RDF(S) definitions into enumerated and dependent types.
To each RDFS class, we assign a corresponding enumerative type. 
We then use these new types as the domain and range of new dependent types, one for each RDF property with the same domain and range.
We may then populate these types with the appropriate RDF edges in a fairly straightforward manner.
Algorithm \ref{algo:conversion} describes this process in more detail.

\begin{algorithm}[H]
\caption{Conversion of knowledge graph into \dtkg{}}
\label{algo:conversion}
\begin{algorithmic}[1]
\FOR {\Co{(subj, pred, obj)} in $\mathcal{KG}$}
\IF {\Co{pred} is \Co{<rdf:type>} and \Co{obj} is not \Co{<rdfs:Class>} or \Co{<rdf:Property>}}
\STATE Construct entity \Co{subj} of enumerated type \Co{Obj}, i.e. \Co{subj : Obj}.
\ENDIF
\ENDFOR

\FOR {\Co{(subj, pred, obj)} in $\mathcal{KG}$}
\IF {\Co{(pred, rdf:type, rdf:Property)} in $\mathcal{KG}$}
\STATE Construct edge \Co{e} of dependent type \Co{Pred subj obj}, i.e. \Co{e : Pred subj obj}
\ENDIF
\ENDFOR

\end{algorithmic}
\end{algorithm}

Note that this algorithm will succeed only if each RDF entity is assigned a class, and all edges obey the RDFS class specifications.
Thus, running this algorithm can be a way of checking that an RDF graph obeys its RDFS domain and range specifications.

\section{Querying of knowledge graphs}\label{sec:queryKG}

Querying is the process of retrieving the information stored in a knowledge graph. 
For RDF graphs, this may be carried out using the SPARQL Protocol And RDF Query Language (SPARQL).
In this section, we demonstrate how to use built-in features of the Coq proof assistant to query \dtkg{}s, and compare these to the equivalent SPARQL queries for RDF graphs.

We introduce a \emph{queries-as-types} approach to querying knowledge graphs (by analogy with the propositions-as-types interpretation of logic given by the Curry-Howard correspondence \cite{Howard1969TheFN}).
In this approach, the process of querying a DTKG involves two steps:
\begin{enumerate}
	\item Define the query as a \emph{record type} \Co{Q};
	\item Answer the query by constructing terms of \Co{Q}.
\end{enumerate}
The record type \Co{Q} identifies a type \Co{T} (e.g. \Co{Person}) in which to search for an answer to the query, and specifies the requirements of the query.
Terms of \Co{Q} may be treated as terms of \Co{T} that satisfy the query, and are thus the answers to the query.
These terms can be constructed in Coq through the use of built-in \emph{tactics}.

In the rest of this section, we elaborate on the queries-as-types approach, and demonstrate how to carry out direct queries and composite queries in DTKGs.



\subsection{Direct queries using \Co{Search}}
\emph{Direct queries} are queries about a single relation or edge in a knowledge graph, such as querying for Sasha's father in Fig \ref{fig:obamagraph-simple}.
In an RDF graph, this is performed using the following SPARQL sequence:

\begin{sparqlin}
SELECT ?person 
WHERE { ex:Sasha ex:father ?person. }
\end{sparqlin}
\begin{sparqlout}
ex:Barack
\end{sparqlout}
The variable of interest \Co{?person} is indicated via the \Co{SELECT} command, followed by a \Co{WHERE} clause that contains the RDF triple pattern to be matched.

Coq's \Co{Search} command is most similar to SPARQL's \Co{SELECT} \dots \Co{WHERE} pattern, and is useful for once-off queries:

\begin{coqin}
Search (FatherOf sasha ?person).
\end{coqin}
\begin{coqout}
witness_fsb: FatherOf sasha barack
\end{coqout}
The output lists all terms that match the search pattern, in this case returning the term \Co{witness_fsb} whose type \Co{FatherOf sasha barack} contains our desired answer \Co{barack}.

However, the results of this query cannot be manipulated in any way.
They cannot be decomposed (for example, to extract just the answer \Co{barack}), they are not stored, and they may not be used as inputs to future queries.
To be able to manipulate query results in these ways, we need to treat queries as types.

\subsection{Direct queries as record types}\label{ss:queriesasrecords}
We now describe our \emph{queries-as-types} approach to querying DTKGs.
Continuing the example above, we wish to query for Sasha's father in the DTKG in Fig \ref{fig:obamacoq}.
We start by defining a record type for the query:

\begin{coqinonly}
Record FatherSasha := 
	{ father : Person; proof_father : FatherOf sasha father }.
\end{coqinonly}
Record types are used in Coq to define tuples with fields that may belong to different types.
Our query types will be record types containing two fields: 
\begin{enumerate}
	\item The first field specifies the type to search over for our answer -- in this case \Co{Person}; 
	\item The second field encodes the query pattern that terms of the first field have to satisfy. In this case, we restrict to those terms \Co{father} in the first field such that \Co{FatherOf sasha father} is inhabited.
\end{enumerate}
Terms of \Co{FatherSasha} are pairs of the form (\Co{father}, \Co{proof_father}) where $\Co{father}$ is of type \Co{Person}, and \Co{proof_father} is a term of \Co{FatherOf sasha father} i.e. a \emph{proof} that \Co{FatherOf sasha father} is inhabited. 
Thus, terms of \Co{FatherSasha} are the answers to our query. Conversely, answering the query involves constructing these terms.

Every record type comes with a \emph{constructor} for creating terms of that type from terms of its fields.
The constructor is called \Co{Build_}\emph{[name of type]} by default.
In our example, the constructor is \Co{Build_FatherSasha}, and we may use it to construct a term of type \Co{FatherSasha}:

\begin{coqin}
Check Build_FatherSasha barack witness_fsb.
\end{coqin}
\begin{coqout}
{| father := barack; proof_father := witness_fsb |}
     : FatherSasha
\end{coqout}
But this assumes that we have already found the terms \Co{barack} and \Co{witness_fsb}, which is the purpose of the query in the first place.

What we want is a method for \emph{finding} the terms \Co{barack} and \Co{witness_fsb} with which to build a term of \Co{FatherSasha}.
To do this, we first define a theorem:

\begin{coqin}
Theorem sasha_father : FatherSasha.
\end{coqin}
This tells Coq that we seek a term of type \Co{FatherSasha}, and that this term will be called \Co{sasha_father} when found.
Our goal is now to prove this theorem (i.e. to produce a term of this type).
Coq contains built-in \emph{tactics} that perform backward reasoning to reduce a goal to smaller subgoal(s).
Since the only way to produce terms of the record type \Co{FatherSasha} is through \Co{Build_FatherSasha}, we call the tactic \Co{eapply} on this constructor.

\begin{coqin}
Theorem sasha_father: FatherSasha.
Proof. eapply Build_FatherSasha.
\end{coqin}
\begin{coqout}
1 subgoal
______________________________________(1/1)
FatherOf sasha ?father
\end{coqout}
Coq reports that our new goal is to find a term of type \Co{FatherOf sasha ?father}, for some existential \Co{?father}. 
Again, we do not wish to supply this term ourselves: we use the \Co{constructor} tactic to tell Coq to look for terms that will resolve this subgoal.
Coq finds such a term and reports that there are no more subgoals:

\begin{coqin}
Theorem sasha_father : FatherSasha.
Proof. eapply Build_FatherSasha. constructor. Defined.
\end{coqin}
\begin{coqout}
No more subgoals.
\end{coqout}
The command \Co{Defined} indicates a completed proof. The proof is given the name \Co{sasha_father} that we included in the theorem statement. 
Printing this proof yields the answer \Co{barack} as well as the term \Co{witness_fsb} found by the \Co{constructor} tactic:

\begin{coqin}
Print sasha_father.
\end{coqin}
\begin{coqout}
sasha_father =  {| father := barack; proof_father := witness_fsb |}
     : FatherSasha
\end{coqout}
Thus, in addition to the query result, query types also provide \emph{explainability} of the result in the form of witnesses.

Finally, if we wish to extract just the answer \Co{barack}, we may project the term \Co{sasha_father} onto its first field using the field name \Co{father}:

\begin{coqin}
Eval compute in (father sasha_father).
\end{coqin}
\begin{coqout}
= barack : Person
\end{coqout}

In this example, there was exactly one inhabitant in the \Co{FatherSasha} query type. In general, there may be no, one or multiple inhabitant(s) to a query type, depending on the availability of the witnesses present in the DTKG.

\subsection{Automation, coercion and abstraction}
The preceding section described the basic method of using record types to create and answer queries.
In this section, we extend the basic method to include additional features.

\bigskip
\noindent\textbf{Automation.} Since all direct queries may be resolved using the \Co{eapply} and \Co{constructor} tactics mentioned above, we combine them into a single tactic that can be applied to automatically construct answers for direct queries.
The tactic language $\mathsf{L_{tac}}$ \cite{Delahaye:2000:TLS:1765236.1765246} lets us create a custom tactic which we call \Co{dqt}\footnote{For `DTKG Query Tactic'.}:

\begin{coqinonly}
Ltac dqt c := eapply c; constructor.
\end{coqinonly}
With this tactic, the proof of the theorem \Co{sasha_father : FatherSasha} may be shortened to:

\begin{coqinonly}
Proof. dqt Build_FatherSasha. Defined.
\end{coqinonly}

\bigskip
\noindent\textbf{Coercion.} 
There may be occasions where we wish to treat terms of our query type as terms of another type.
For instance, it might make sense to treat terms of \Co{FatherSasha} as terms of \Co{Person}.
Coq allows us to do this via coercion of types, by modifying the declaration of our query type:

\begin{coqinonly}
Record FatherSasha := 
	{ father :> Person; proof_father : FatherOf sasha father }.
\end{coqinonly}
We have replaced the field \Co{father : Person} with \Co{father :> Person}, which tells Coq to coerce terms of type \Co{FatherSasha} into terms of type \Co{Person} when required. 
For instance, this lets us carry out an equality comparison\footnote{The \Co{eqType} in Mathematical Components \cite{mahboubi2017mathematical} is enabled here for the use of \Co{==}.} between \Co{barack} and \lstinline{sasha_father}: 

\begin{coqin}
Eval compute in barack == sasha_father.
\end{coqin}
\begin{coqout}
= true : bool
\end{coqout}
This also works on results of different queries (assuming \Co{malia_father} has been similarly defined as a term of \Co{FatherMalia}):

\begin{coqin}
Eval compute in (sasha_father : Person) == malia_father.
\end{coqin}
\begin{coqout}
= true : bool
\end{coqout}
For this comparison to work, we needed to coerce \Co{malia_father} to be of type \Co{Person}.
Evaluating \Co{sasha_father == malia_father} would throw an error, as these are terms of different types.

\bigskip
\noindent\textbf{Abstraction.}
Finally, instead of creating query types \Co{FatherSasha}, \Co{FatherMalia} and so on, we may wish to abstract these query types into a pattern that may be applied to any term of \Co{Person}.
This may be implemented via a dependent record type of the form:

\begin{coqinonly}
Record Father (x : Person) :=
  { father :> Person; proof_father : FatherOf x father }.
\end{coqinonly}
We can then use the query types \Co{Father sasha}, \Co{Father malia} and so on, without having to define a new query type for each \Co{Person}.
The query type \Co{Father sasha} may be used in the same way as \Co{FatherSasha}:

\begin{coqin}
Theorem sasha_father : Father sasha.
Proof. dqt Build_Father. Defined.

Print sasha_father.
\end{coqin}
\begin{coqout}
sasha_father =  {| father := barack; proof_father := witness_fsb |}
     : Father sasha
\end{coqout}
Note that the same proof \Co{dqt Build_Father} may be used to produce terms of \Co{Father sasha}, \Co{Father malia}, and so on.

More than just a convenience, abstracting direct queries in this manner is essential for defining more complex queries, such as the composite queries that we now turn to.

\subsection{Composite queries}
\label{ss:compq}
The knowledge graph of Fig \ref{fig:obamagraph-simple} does not directly show that Malia's father is Barack.
However, if we had additional information of the form, ``\emph{The husband of Malia's mother is Malia's father}'', we can deduce that Malia's father is Barack from the fact that Malia's mother is Michelle, and Michelle's husband is Barack.

Deductions of this form may be carried out in knowledge graphs by using \emph{composite queries}, which are queries whose answers require information from multiple edges. 
This is done in SPARQL as follows:

\begin{sparqlin}
SELECT ?father
WHERE { ex:Malia ex:mother ?mother;
           ?mother ex:husband ?father. }
\end{sparqlin}
\begin{sparqlout}
ex:Barack
\end{sparqlout}

We demonstrate how similar queries may be defined for \dtkg{}s using record types. 
We first assume that we have constructed the following direct queries:

\begin{coqinonly}
Record Mother (x : Person) :=  
  { mother :> Person; proof_mother : MotherOf x mother }.
Record Husband (x : Person) := 
  { husband :> Person; proof_husband : HusbandOf x husband}.
\end{coqinonly}
We define a new  record type \Co{Father'} whose fields encapsulate the logic that the husband of the mother is the father:

\begin{coqinonly}
Record Father' (x : Person) := 
  {  mother' : Mother x; father' :> Husband mother' }.
\end{coqinonly}
The first field contain the mother of \Co{x}, while the second field contains the husband of the mother in the first field.
Since our desired answer is in the second field, we apply coercion there instead of in the first field.

Malia's father, although not directly represented in the knowledge graph, may be inferred using the query \Co{Father' malia}:

\begin{coqin}
Theorem malia_father: Father' malia.
Proof. unshelve eapply Build_Father'.
\end{coqin}
\begin{coqout}
2 subgoals
______________________________________(1/2)
Mother malia
______________________________________(2/2)
Husband ?mother'
\end{coqout}
As before, we call \Co{eapply} on the constructor \Co{Build_Father'} to reduce this goal to its subgoals.
The tactic \Co{unshelve} moves all subgoals into focus (otherwise we can only see the final goal).

The first subgoal is to produce a term of type \Co{Mother malia}. This is a direct query, and can be resolved with our custom \Co{dqt} tactic:

\begin{coqin}
Theorem malia_father: Father' malia.
Proof. unshelve eapply Build_Father'. 
dqt Build_Mother.
\end{coqin}
\begin{coqout}
1 subgoal
______________________________________(1/1)
Husband {| mother := michelle; proof_mother := witness_mmm |}
\end{coqout}
Resolving the first subgoal updates the second subgoal: \Co{?mother} is instantiated with the term created from querying Malia's mother. 
Thanks to coercion, we may treat this subgoal as one of producing a term of \Co{Husband michelle}, which is again a direct query that may be resolved with \Co{dqt}:

\begin{coqin}
Theorem malia_father: Father' malia.
Proof. unshelve eapply Build_Father'. 
dqt Build_Mother. 
dqt Build_Husband. Defined.
\end{coqin}
\begin{coqout}
No more subgoals.
\end{coqout}

\begin{figure}
\centering
\begin{tikzpicture}[scale = 1, node distance=\layersep,>=stealth]
    \tikzstyle{neuron}=[circle,fill=black!25,minimum size=17pt,inner sep=0pt];
    \tikzstyle{unit}=[neuron, fill=red!50,thick];
    \tikzstyle{babi}=[rectangle,fill=blueTOL,inner sep=3pt,minimum size=10pt,,rounded corners=1mm,
    ];
        \tikzstyle{loc}=[rectangle,fill=black,inner sep=3pt,minimum size=10pt,,rounded corners=1mm];
\node[babi,draw=white,thick] (f) at (-1.75,0) {{\color{white}\textsf{\scriptsize barack}}};
\node[babi,draw=white,thick] (d) at (5.25,0) {{\color{white}\textsf{\scriptsize malia}}};
\node[babi,draw=white,thick] (s) at (-4.75,0) {{\color{white}\textsf{\scriptsize sasha}}};
\node[babi,draw=white,thick] (m) at (1.75,0) {{\color{white}\textsf{\scriptsize michelle}}};

\path (s) edge [->, thick,]  node[align=center,sloped] {\scriptsize$\mathsf{witness\_fsb}$\\ \scriptsize$\mathsf{: FatherOf~s~b}$} (f);

\path (d) edge [->,  thick]  node[sloped, align=center] {\scriptsize$\mathsf{witness\_mmm}$ \\ \scriptsize$\mathsf{: MotherOf~m~m~~~}$} (m);
\path (m) edge [->,  thick]  node[align=center,sloped] {\scriptsize$\mathsf{witness\_hmb}$\\ \scriptsize$\mathsf{: HusbandOf~m~b}$} (f);

\path (d) edge [->,  thick, bend right=30, dashed]  node[align=center,sloped] {\scriptsize$\mathsf{malia\_father}$\\ \scriptsize$\mathsf{: Father'~malia}$} (f);

\end{tikzpicture}

%
%
%
\caption{Composite queries infer new information by constructing new (dashed) edges. (We have used \Co{FatherOf s b} as shorthand for \Co{FatherOf sasha barack} and so on.) }
\label{fig:typedobamafam}
\end{figure}
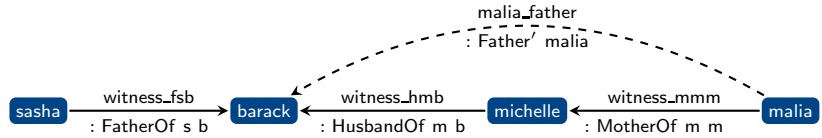

The constructed term \Co{malia_father} can be interpreted as a directed edge from \Co{malia} to \Co{barack} composed from the edges \Co{witness_mmm} and \Co{witness_hmb}, as illustrated in Fig \ref{fig:typedobamafam}.
It is coerced to \Co{barack} when passed to functions or dependent record types expecting an argument of type \Co{Person}:

\begin{coqin}
Eval compute in barack == malia_father.
\end{coqin}
\begin{coqout}
= true : bool
\end{coqout}

In this section, we have seen how record types may be used to define both direct and composite queries, which may be answered using our custom \Co{dqt} tactic.
We have also noted the key role of coercion and abstraction in building up composite queries from direct ones.
We are optimistic that Coq can allow the creation and answering of even more complicated queries on \dtkg{}s.

\subsection{Features of \dtkg{}s}
\label{ss:featuresdtkg}
We close this section by highlighting features of \dtkg{}s which are not present in RDF graphs.

\bigskip
\noindent\textbf{Compositionality.}
The abstraction of the query types like \Co{FatherSasha} to the dependent record types like \Co{Father (x : Person)} allows queries of the same intent in different contexts (represented by the argument \Co{x : Person} in this case). 
In particular, it allows answers from earlier queries to be used as arguments for future queries.
This is crucial in making direct queries the building blocks of more complicated composite queries, such as those in Section \ref{ss:compq}.
More generally, we can build more complex queries using any type-compatible combination of direct and composite queries.

\bigskip
\noindent\textbf{Explainability.}
The results returned from querying \dtkg{}s contain: (1) the answer to the query, and (2) witnesses supporting the query. 
At the \dtkg{} level, witnesses are atomic objects that provide explanations for answers to direct queries.
For composite queries,  explanations are supported by the multiple witnesses required to make the neccessary logical deductions that leads to the query result. Explanations for all queries are purposefully incorporated in our answers, and implemented by record types (which generalize dependent pair types). 
The use of tactics automates the construction of both the answers and the witnesses to queries.

\bigskip
\noindent\textbf{Proof relevance.} 
Combining queries-as-types with propsitions-as-types allows us to view query answers and witnesses as proofs of propositions. 
As mentioned in \ref{ss:adv-dtkg-rdf}, \dtkg{}s may contain multiple edges for the same relation between two entities, such as \Co{witness_BC} and \Co{witness_DNA} for \Co{FatherOf sasha barack}. Thus, answers to queries may have multiple witnesses to their veracity. Proof relevance is a concept in constructive mathematics which treats mathematical proofs as objects and places importance on how a proof was constructed \cite{nlab:proof_relevance}. Adopting proof relevance provides a clear distinction between answers constructed from different witnesses. 
For instance, the terms \Co{witness_BC} and \Co{witness_DNA} lead to different query results \Co{sasha_father_BC} and \Co{sasha_father_DNA} which are not equal 
(\lstinline{sasha_father_BC = sasha_father_DNA} cannot be proven)
even though they may both coerce to the same term:

\begin{coqin}
Eval compute in (sasha_father_BC : Person) == sasha_father_DNA.
\end{coqin}
\begin{coqout}
= true : bool
\end{coqout}

\section{Conclusion}\label{sec:conclusion}

In this paper, we have demonstrated how to represent and query knowledge graphs in dependent type theory. 
More than just reproducing the functionalities of RDF graphs and SPARQL queries, we have seen how our DTKGs make use of dependent types and tactics in Coq to provide explainability of query results through the construction of witnesses in a compositional and automated manner.

Although we only treat the encoding of RDF graphs and SPARQL queries in this paper, we believe that dependent type theory is sufficiently expressive to represent constructs from OWL, RDFS and RIF as well.
Future work will reproduce these constructs for carrying out more sophisticated reasoning on DTKGs.
This paper is just a first step towards our larger goal of reasoning with knowledge graphs in dependent type theory, and we hope that it will serve as a proof-of-concept and an impetus for future efforts in this direction.




%
%
%
\bibliographystyle{splncs04}
\bibliography{dtkg}
%
%
%
%
%
\end{document}